\documentclass[9pt, a4paper, one column]{article}
 \pdfoutput=1
\usepackage{geometry}
\usepackage[affil-it]{authblk}
\usepackage[english]{babel}
\usepackage[comma, super]{natbib}
\usepackage[usenames, dvipsnames]{color}
\usepackage[colorinlistoftodos]{todonotes}
\usepackage[font=small]{caption}
\usepackage{subcaption}
\usepackage{amsmath}    
\usepackage{graphicx} 
\usepackage[T2A,T1]{fontenc}
\DeclareSymbolFont{cyrillic}{T2A}{cmr}{m}{n}
\DeclareMathSymbol{\Comb}{\mathalpha}{cyrillic}{217}
\DeclareMathSymbol{\comb}{\mathalpha}{cyrillic}{216}
\usepackage[utf8]{inputenc}
\usepackage{csquotes}
\begin{document}

\title{Frequency-time coherence for all-optical sampling without optical pulse source}
\author{S.~Preussler}
\author{G.~Raoof Mehrpoor}
\author[*]{T.~Schneider}

\affil{Insitut für Hochfrequenztechnik, Technische Universität Braunschweig,\par
Schleinitzstr. 22, 38106 Braunschweig, Germany}
\affil[*]{Corresponding author: thomas.schneider@ihf.tu-bs.de}
\date{\vspace{-5ex}}

\maketitle 

\textbf{Sampling is the first step to convert an analogue optical signal into a digital electrical signal. The latter can be further processed and analysed by well-known electrical signal processing methods. Optical pulse sources like mode-locked lasers are commonly incorporated for all-optical sampling, but have several drawbacks. A novel approach for a simple all-optical sampling is to utilise the frequency-time coherence of each signal. The method is based on only using two coupled modulators driven with an electrical sine wave, allowing simple integration in appropriate platforms, such as Silicon Photonics. The presented method grants all-optical sampling with electrically tunable bandwidth, repetition rate and time shift.}\par
\subsection*{Introduction}
Analog-to-digital conversion (ADC) is essential for precise transformation and processing of information carried by electromagnetic signals into their binary correspondent. In computer technology, the powerful software improvement has kept pace with more complex hardware development. This is not the case in electronic systems, since the ADC and sampling rate of signals is a limiting factor for the performance. By the rapid growth of photonic technologies, there has been a great interest in reconciling the ADC requirement with optical components during the last 25 years. The electronic ADC schemes are limited by several noise sources such as quantisation noise, thermal noise and nonlinearities \cite{ADC_Rev, MW_Theory},
motivating further experimental investigations on photonic counterpart.\par
According to the well-known Nyquist sampling theorem, to sample and restore a signal with maximum frequency of $f_{max}$ Hz properly, pulses with a minimum repetition rate of $2f_{max}$ should be utilised. This complies a sampling frequency of $f_s=2f_{max}=1/t_s$, where $t_s$ is the time interval between two sampling points.
To achieve high resolution optical sampling, different methods have been proposed.
The simplest method for sampling of an optical signal is to utilise a fast photodetector in combination with sample-and-hold circuits. However, the signal is sampled completely in the electrical domain and therefore restricted by the bandwidth of the electronic components. Another approach is based on a multiplication between the signal and a train of short laser pulses (see the Methods for details). The multiplication can be carried out in a modulator, driven by the electrical signal \cite{ADC_Rev}. For this method the sampling rate is restricted by the bandwidth of the modulator. Faster sampling is achieved by multiplication in a nonlinear element \cite{Monitoring, Eye_D}. A temporal magnifier such as a time lens can be used for very short optical waveforms \cite{High_S}. All mentioned techniques require a stable short-period optical pulse source.\par
Conventionally, a mode-locked laser (MLL) is utilised as the pulse source to perform optical sampling, where the sampling rate depends on the repetition rate of the MLL. Besides the cost and limited stability of MLLs, they are not flexible to integrate into an electronic system. The amplitude noise and time jitter between the pulses are remarkable, affecting the ADC accuracy. Additionally, the repetition rate of the pulses and therefore the sampling frequency depends on the optical path length of the MLLs cavity. A tuning of the sampling frequency is therefore rather complicated and only possible in a limited range. Furthermore, for a parallelisation of the sampling in time, tunable delay lines for each branch and an exact calibration is required. For a parallelisation in the frequency domain, interleaved pulses of different wavelengths are necessary \cite{150G}.\par
In this paper, we demonstrate a simple all-optical sampling, without requiring any pulse source. All sampling parameters can be simply and fast tuned in the electrical domain and a parallelisation of sampling in time and frequency domain is straight forward.
\subsection*{Theory}
The method is based on a convolution between a frequency comb and the signal spectrum in two coupled modulators. A similar technique can be used for the demultiplexing of Nyquist channels \cite{MAS}.
\begin{figure}[ht]
\centering
\includegraphics[width=0.8\linewidth]{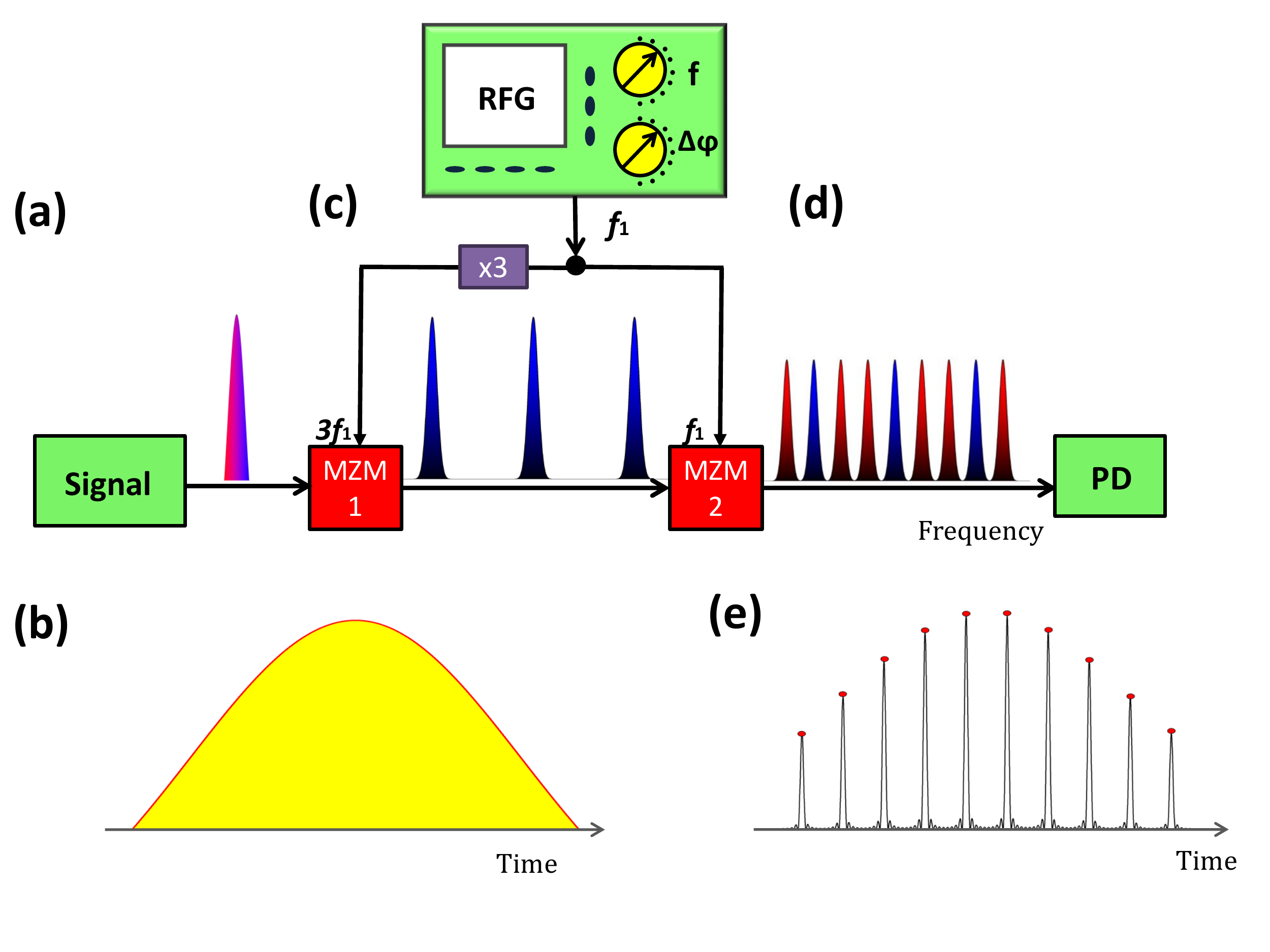}
\caption{Basic principle of optical sampling based on frequency-time coherence. The insets (a) and (b) show the input signal in the frequency and time domain, respectively. In the first Mach-Zehnder modulator (MZM1) the spectrum of the input signal is tripled (inset (c)). The second modulator leads to the spectral components shown in red in inset (d). In order to adapt the sampling parameters to the signal, the frequency as well as the phase of the sinusoidal wave generated by the radio frequency generator (RFG) can be changed. The result is a convolution of the input spectrum with a nine-line rectangular frequency comb, corresponding to an optical sampling with a Nyquist pulse sequence in the time domain (inset e). The sampling point is the integration of the sequence in the time-interval of the repetition rate, which can be carried out with a low bandwidth photodiode (PD).}\label{fig:shapes}
\end{figure}
The basic principle of all-optical sampling based on frequency-time coherence is depicted in Fig.~\ref{fig:shapes}. For this sampling just two coupled optical Mach-Zehnder modulators (MZM) are required. The sinusoidal input frequency $f_1$ is generated by a radio frequency (RF) generator and can be changed in frequency and phase. This input frequency is tripled in a frequency tripler and used to drive the first modulator (MZM1). The second modulator (MZM2) is directly driven with $f_1$. The bias voltage and RF power of both modulators are adjusted such that just two new sidebands with equal power and phase as the carrier are generated \cite{Nature}. The lengths of the electrical and optical connections have to be compensated, which can be done by an additional phase shifter either in the $f_1$ or $3f_1$ branch (not shown). 
For the sake of simplicity, it is assumed that the optical input signal has a band limited spectrum with a Gaussian-like shape and a frequency and time representation shown in Fig.~\ref{fig:shapes}a and Fig.~\ref{fig:shapes}b, respectively. 
The first modulator simply triples the spectrum as can be seen from Fig.~\ref{fig:shapes}c. The frequency separation between the spectra is defined by $3f_1$. The second modulator leads again to two new copies of the input spectrum around each carrier. Since three input carriers are present, the results are the six red copies of the input spectrum, shown in Fig.~\ref{fig:shapes}d.
Thus, the output spectrum consists of nine equal copies of the input spectrum, with a frequency separation of $f_1$. Since the modulators were adjusted in a way that the sidebands have the same amplitude as the carrier, the nine copies are almost equal to each other. The obtained output spectrum can be seen as a convolution between a nine-line rectangular frequency comb and the input spectrum. This convolution corresponds to a multiplication between a sinc-pulse sequence and the time signal in the time domain, as can be seen in Fig.~\ref{fig:shapes}e.
Thus, the result is an optical sampling of the time domain signal by a sinc-pulse sequence. The same can be expressed mathematically by the equation:
\begin{eqnarray}
\label{D1}
\mathcal{F}^{-1}\Big\{X(f)\ast\big[\Comb(f)\times\Pi(f)\big]\Big\}\nonumber\\
=x(t)\times(\comb(t)\ast\text{sinc}(t)),
\end{eqnarray}
where $\mathcal{F}^{-1}$ is the inverse Fourier transform and $\ast$ depicts the convolution. 
Here $\comb(t)$ is the time domain representation of an unlimited frequency comb, $\Comb(f)$, 
\begin{equation}\label{Comb(f)}
\Comb(f)=\sum_{n=-\infty}^{\infty}\delta(f_c-nf_1),
\end{equation}
and the rectangular function $\Pi(f)$ is defined along the bandwidth of the comb in frequency domain 
\begin{equation}
\label{Pi(f)}
\Pi(f)=
\begin{cases}
1 & -\frac{Nf_1}{2}\le f_c\le\frac{Nf_1}{2}\\
0 & \text{otherwise}.
\end{cases}
\end{equation}
$f_1$ denotes the frequency spacing between the comb lines, with $N$ the total number of lines, determining $B=Nf_1$ as the required bandwidth for the rectangle and $f_c$ as the carrier frequency. This definition leads to an odd number of lines in the comb. For an even number of comb lines, the time domain representation is a sinc-pulse sequence as well \cite{Nature}.\par
If the input signal is a Dirac delta function, i.e. $X(f)=\delta(f)$, the time response would be $x(t)=1$. Therefore, if the input to the two modulators is a narrow linewidth optical wave, the output would be a rectangular nine-line frequency comb, or a sinc-pulse sequence in time domain. The repetition time of the sequence corresponds to the inverse of the frequency spacing between the lines, or the RF frequency applied to the modulator $t_s=1/f_1$. Accordingly, the inverse Fourier transform of a multiplication between a frequency comb and a rectangular function leads to a sinc-pulse sequence \cite{Nature, Soto, Flexible}:
\begin{equation}
\label{f-1}
\mathcal{F}^{-1}\Big\{\Comb(f)\times\Pi(f)\Big\}=\comb(t)\ast\text{sinc}(t).
\end{equation}
%
Due to reciprocity of the Fourier transform, a multiplication in frequency corresponds to a convolution in time domain. Thus, the right hand side of equation~\eqref{f-1} is the convolution of the two functions in time domain. Having the function definitions in time domain as
\begin{align}
\comb(t)&=\sum_{n=-\infty}^{\infty}\delta(t-n/f_1),\\
\text{sinc}(t)&=\dfrac{\sin(\pi t)}{\pi t},
\end{align}
the convolution can be written as \cite{Nature}: 
\begin{equation}
\comb(t)\ast \text{sinc}(t)=\dfrac{\sin(\pi Nf_1t)}{N\sin(\pi f_1t)}.
\end{equation}
Generally, the resulting Nyquist sequence of equation~\eqref{f-1} is \cite{Nature}
\begin{equation}
\label{Nyquist}
\dfrac{\sin(\pi Nf_1t)}{N\sin(\pi f_1t)}=\sum_{n=-\infty}^{\infty}(-1)^{(N-1)n}\text{sinc}\big(\pi Nf_1\big(t-\dfrac{n}{f_1}\big)\big).
\end{equation}
Thus, the desired Nyquist pulse sequence is the unlimited summation of single sinc-pulses, each of which is shifted by $n/f_1$ and unlimited in time domain and therefore non-causal. The interference-free summation of these unlimited pulses is only possible due to the orthogonality of the pulses to each other, if the time shift corresponds to $k/(Nf_1)$ where $k$ is an integer.\par
Suppose an arbitrary signal $X(f)$ in frequency domain with $x(t)$ being the signal in the time domain. The convolution of this signal in the frequency domain with a rectangular frequency comb corresponds to a multiplication between the signal and the Nyquist pulse sequence in time domain:
\begin{eqnarray}
\label{exp}
&\mathcal{F}^{-1}\Big\{X(f)\ast\big[\Comb(f)\times\Pi(f)\big]\Big\} \nonumber\\
&=x(t)\times(\comb(t)\ast\text{sinc}(t))\nonumber \\
&=x(t)\times\dfrac{\sin(\pi Nf_1t)}{N\sin(\pi f_1t)}.
\end{eqnarray}
The sampled signal $x_N(t)$ is achieved by the right-hand side of equation~\eqref{exp}. With the order reversed, the Fourier transform of the sampled signal can be found as
\begin{align}
\label{F-1}
\mathcal{F}\Big\{x_N(t)\Big\}&=\mathcal{F}\Big\{x(t)\times(\comb(t)\ast\text{sinc}(t))\Big\}\nonumber\\
&=X(f)\ast\big[\Comb(f)\times\Pi(f)\big].
\end{align}
Thus, the convolution of the signal spectrum with the rectangular comb through modulation by the two coupled modulators (as shown in Fig.~\ref{fig:shapes}), is equal to a multiplication of an externally generated sinc-pulse sequence by the time domain signal (as in Fig.~\ref{fig:parall.freq}).\par
\subsection*{Experiment}
\begin{figure}[ht!]
\centering
\begin{subfigure}[b]{\linewidth}
\centering
\includegraphics[width=0.6\linewidth]{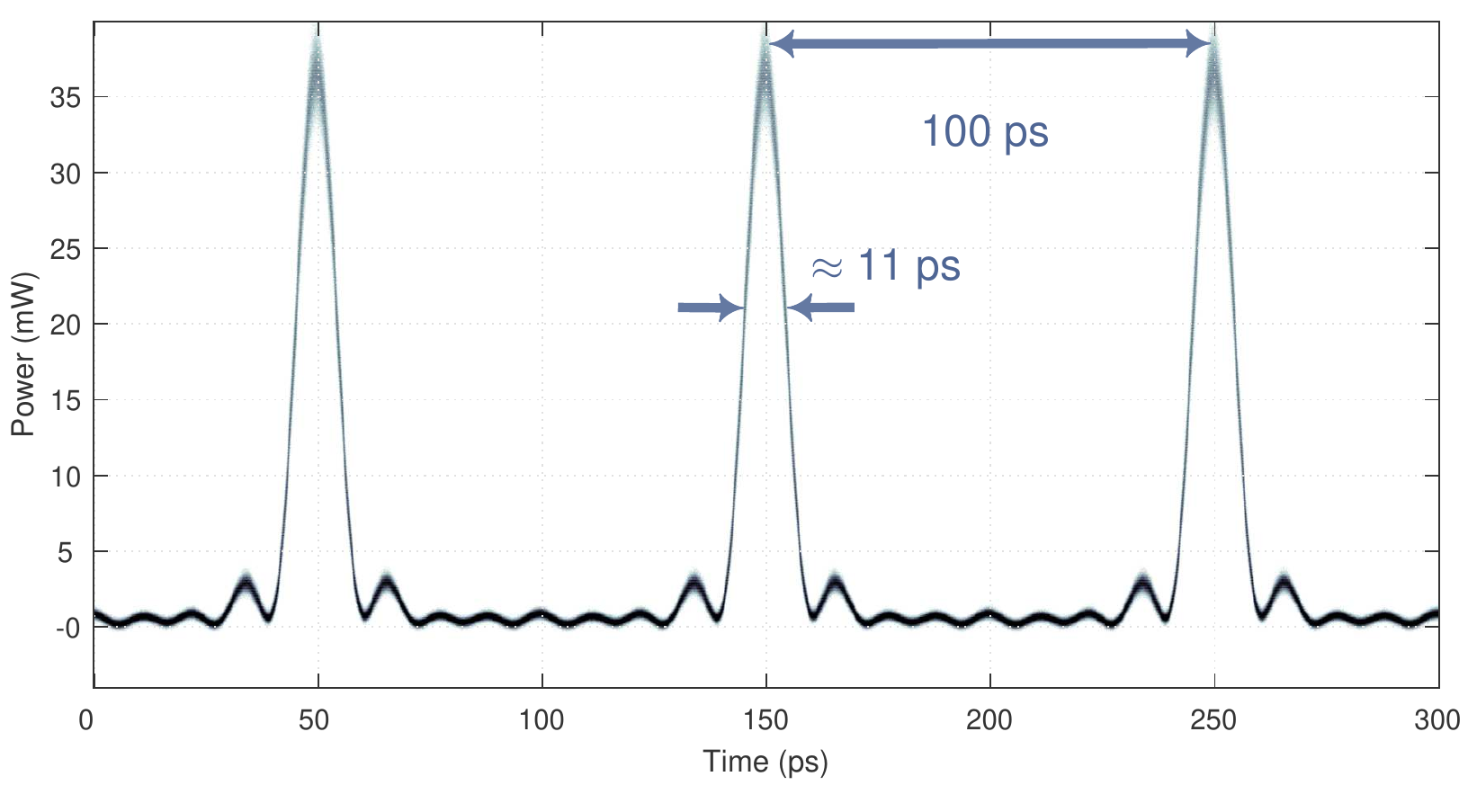}
\caption{}\label{fig:colorgrade}
\end{subfigure}
 \centering
\begin{subfigure}[b]{\linewidth}
\centering
\includegraphics[trim={40 20 20 20},width=0.6\linewidth]{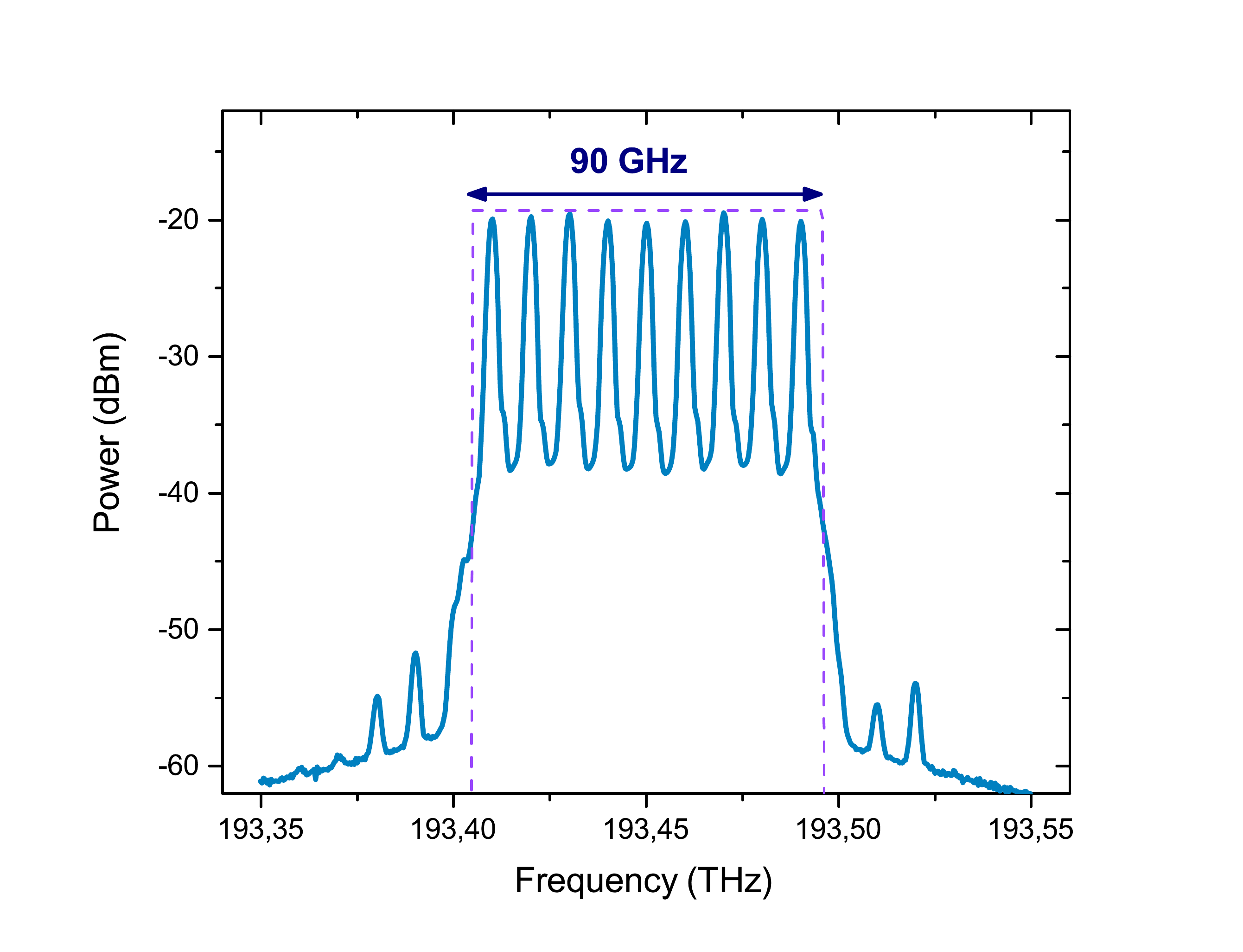}
\caption{}\label{fig:comb}
\end{subfigure}
\caption{(a) Generated sinc-shaped pulse sequence in time domain, if a narrow linewidth optical wave is used as the input to the two modulators; (b) corresponding frequency domain representation of the $9$-line spectral comb with $f_1=10$ GHz. The dashed rectangle is the bandwidth of one
single sinc-pulse with the same duration. The broad width of the lines is a result of the limited resolution of the used optical spectrum analyser. The linewidth is a result of the laser source and in the kHz-range, thus around $6$ orders of magnitude smaller than the comb bandwidth.}\label{fig:Nyquist}
\end{figure}

First, with a CW signal as input, sinc-shaped pulse sequences were generated as shown in Fig.~\ref{fig:colorgrade}.
The according frequency domain representation, i.e. the frequency comb, can be seen in Fig.~\ref{fig:comb}.
The $9$ comb lines have a frequency spacing of $10$ GHz. For a single pulse, this corresponds to a bandwidth of $B=Nf_1=90$ GHz (See the dashed rectangle in Fig.~\ref{fig:comb}). The time to the first zero-crossing is accordingly $\tau=1/B\approx11$ ps which corresponds approximately to the full-width at half-maximum duration of the pulses. 
Since both modulators are driven only with two RF frequencies generated from a single source, the generated sinc-pulse sequence is highly stable and shows a notably low noise-level. As expected, high quality sinc-shaped pulses are achieved with time jitter of less than $82$ fs and low amplitude noise corresponding to a high signal- to-noise ratio level \cite{Nature}.
The bandwidth and repetition time of the pulses can be changed very fast and precisely by an alteration of the RF input frequency. Additionally, the sampling point can be shifted in respect to the signal by a phase change of the electrical signal. Thus, the sampling can be adapted to the signal and it can be scanned through it electrically. Correspondingly, all-optical sampling of an arbitrary signal is realised through the introduced approach with fully tunable sampling rate. Thus, the technique offers flexible sampling of the signals at lower costs, without requiring pulse sources.\par
%
\begin{figure}[ht]
\centering
\includegraphics[width=\textwidth]{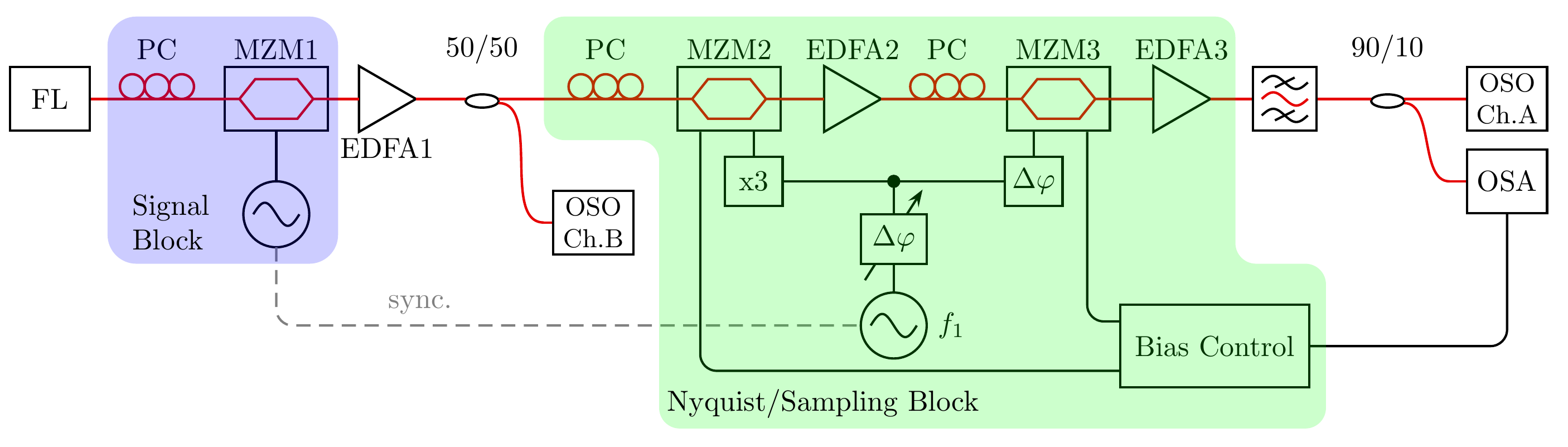}
\caption{Detailed experimental setup. FL: fiber laser, PC: polarization controller, MZM: Mach-Zehnder modulator, EDFA: Erbium doped fibre amplifier, OSO: optical sampling oscilloscope, OSA: optical spectrum analyser.}\label{fig:Setup}
\end{figure}  
Figure~\ref{fig:Setup} indicates the basic experimental setup for the proposed all-optical sampling technique. The optical signal to sample is generated from an electrical signal in the \enquote{Signal Block}. The electrical signal is converted to the optical domain by a conventional Mach-Zehnder modulator (MZM1). A CW fibre laser (FL) operating at the wavelength of $1549.72$ nm with $40$ mW output power generates the optical carrier.
Considering power loss caused by the MZM in the Signal Block the modulated signal is amplified by the first Erbium-doped fibre amplifier (EDFA1) afterwards. The output power of $17$ dBm is split in a $3$-dB coupler in order to monitor the desired optical signal at one channel of an optical sampling oscilloscope (OSO). Subsequently, the signal is fed to the \enquote{Nyquist/Sampling Block} including two coupled MZMs driven by two sinusoidal RFs generated from one generator. 
The output signal of MZM2 is amplified by EDFA2 to $10$ dBm to compensate the power loss during modulation through MZM3. The RF $f_1$ is produced by an RF generator. The pulsewidth as well as the point where the signal is sampled can be controlled by this modulation frequency. The first modulator in the Sampling Block (MZM2) is driven with the tripled frequency $3f_1$. The variable phase shifter is responsible for the time delay of the sinc-pulse train. At a frequency of $10$ GHz it is capable of shifting the phase more than $380$ degrees. Afterwards, the signal is split. In one branch the frequency is tripled and fed to MZM2. In the second branch the electrical path difference is adapted by a phase shifter and fed to MZM3. Additionally, both generators are synchronised with each other.\par
\begin{figure}[!ht]
\centering
\includegraphics[trim={60 0 30 0},width=0.7\linewidth]{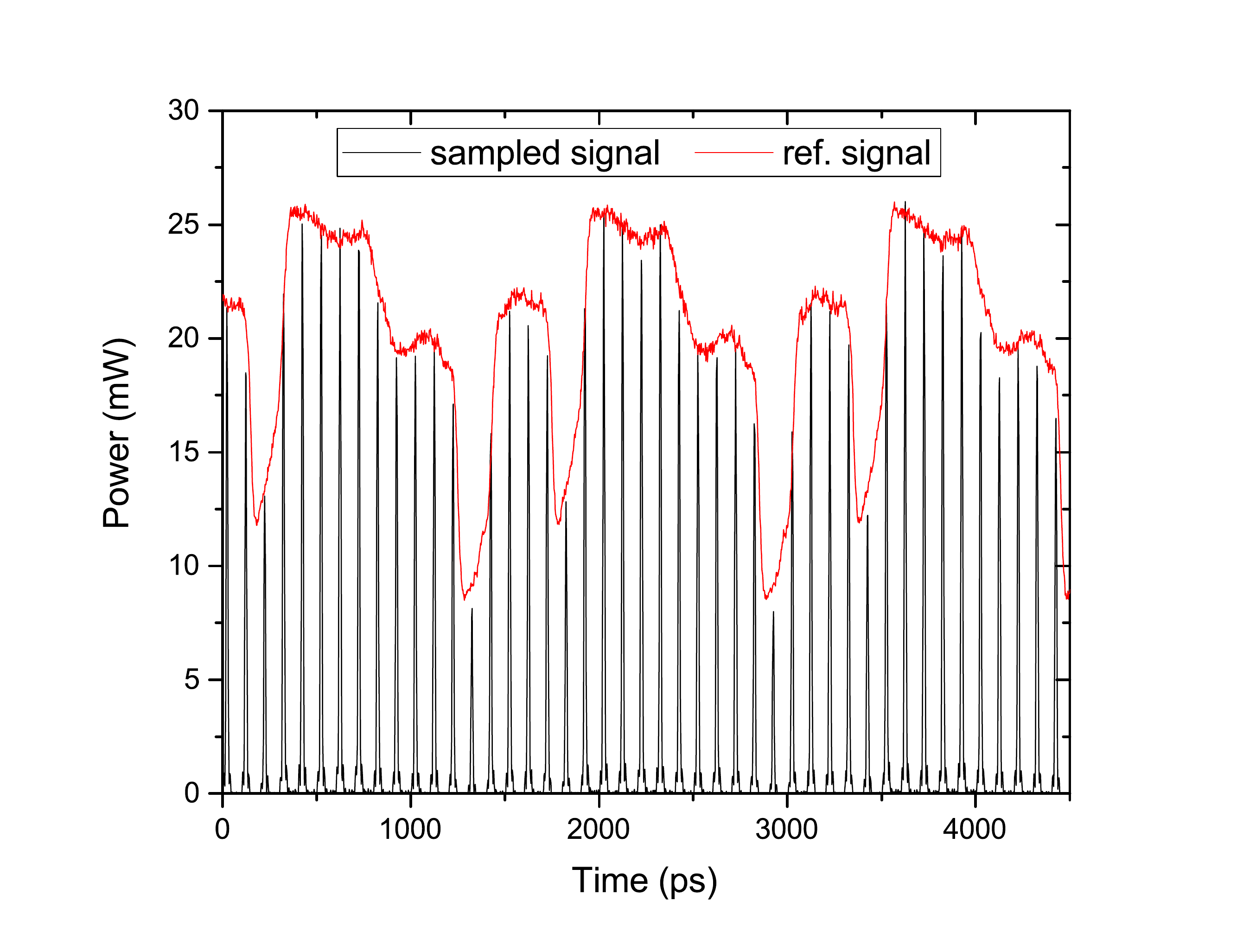}
\caption{Optically sampled 16-bit data packet created by a pattern generator with data rate of $10$ Gbps. The frequency $f_1$, applied to the coupled modulators, was $10$ GHz.}\label{fig:packet}
\end{figure}
\newpage
For the stabilisation of the frequency comb, the bias voltage of the modulators is controlled in a simple feedback loop. Thus, an optical spectrum analyser (OSA) measures the maximum amplitudes of the copied spectra and a computer program adjusts the bias voltage in a way, that all these amplitudes remain equal. Output powers of the amplifiers are adjusted to constant values in order to assure the accuracy of the measurement at each step. All modulators are equipped with polarisation controllers (PC) in order to compensate the polarisation alterations along the fibres and obtain the maximum output power.
The polarisation is aligned for maximum transmission. After the Sampling Block, the signal convolved with the frequency comb is amplified in EDFA3 in order to provide sufficient input power for the OSO. With the help of a $1$ nm band-pass filter the out-of-band noise components are reduced.\par
We have investigated optical sampling of an arbitrary $16$-bit data packet to prove the performance of the proposed technique. A pattern generator was used to create the intended data packet with a repetition rate of $100$ ps. The resulting optically sampled signal can be seen in Fig.~\ref{fig:packet}. For visualisation a commercial OSO was used.
The red curve shows the desired signal, while the black lines correspond to the sampled signal with a rate of $100$ ps. During the measurement the OSO was triggered by the sampled waveform. As can be seen, regardless of the quality of the signal from the pattern generator, it is multiplied by the pulses and sampled, respectively. Conforming to electronics, the all-optical sampling can be accomplished for periodic or non-periodic signals. Since the generated pulse-sequences possess very high quality (see Fig.~\ref{fig:shapes}a), we address the slight differences between the reference and the sampled signal to trigger problems of the OSO.
The point of sampling can be varied by changing the phase of the RF frequency used to drive the modulators. Thus, for periodical signals a simple and fast scanning through the signal is possible by an electrical phase change.\par
\begin{figure}[!ht]
\centering
\includegraphics[trim={50 20 50 20},width=0.7\linewidth]{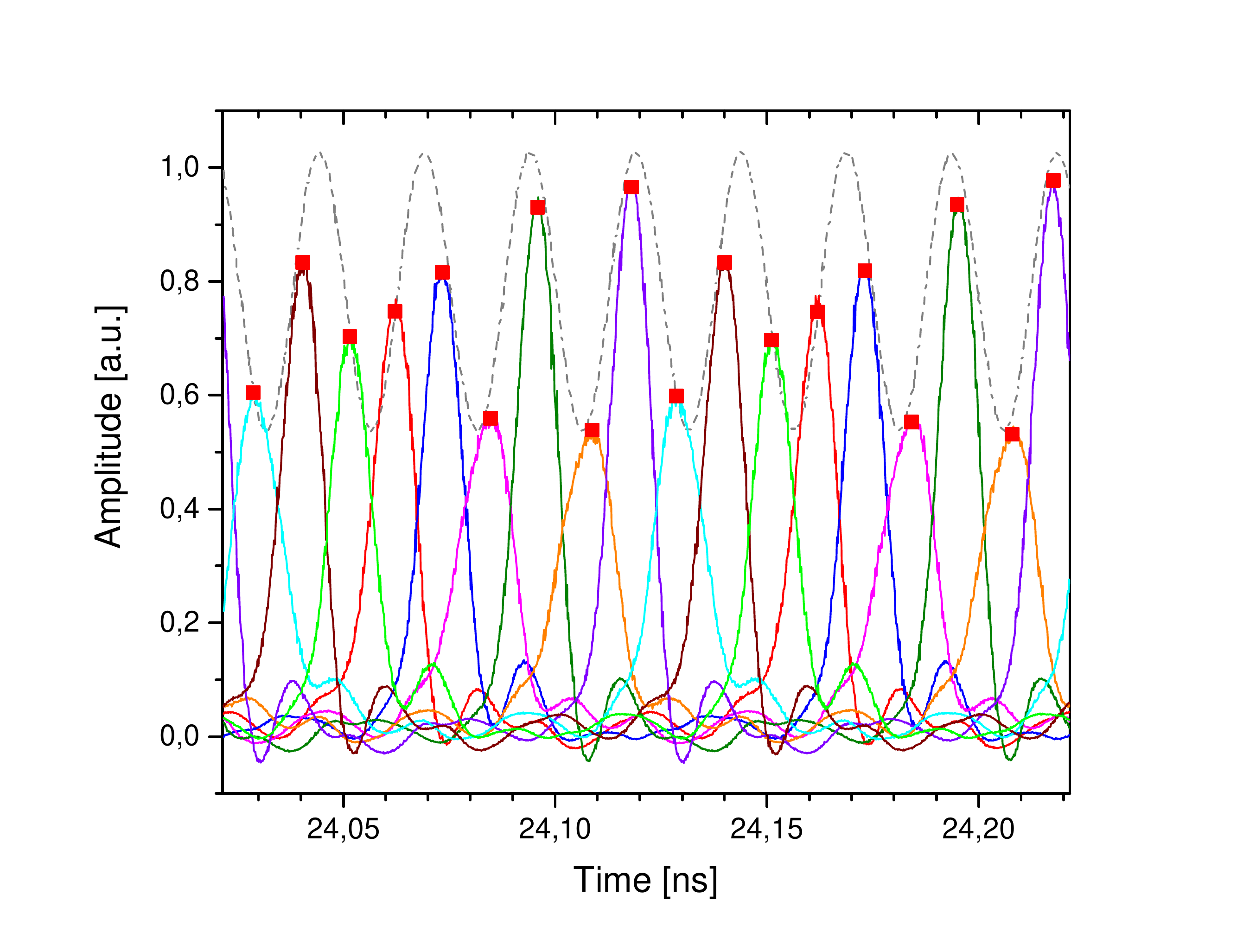}
\caption{Optical sampling of a $40$ GHz sinusoidal signal. The reference is displayed by the gray curve, the coloured traces represent the pulse trains at different phase values and the red dots the integrated, or the sampling values, respectively. The measurement was carried out with low bandwidth equipment.}\label{fig:Graph3}
\end{figure}

The measurement of a $40$ GHz sinusoidal signal can be seen in Fig.~\ref{fig:Graph3}, where the reference trace is represented by the dashed gray line. Therefore, the OSO was replaced by a photodiode and an oscilloscope. Due to the limited bandwidth of the photodiode ($50$ GHz) the pulses are distorted. Within each trace of the measurement two consecutive pulses are shown and analysed. During the measurement the phase of $f_1$ was shifted, leading to a delay of the sinc-pulse train in steps of $11.1$ ps. Finally, the signal traces were recorded and integrated by software within a time window of $100$ ps. The result is illustrated by the red rectangles in Fig.~\ref{fig:Graph3} representing the sampling points.
%
\subsection*{Discussion}
\begin{figure}[!ht]
\centering
\includegraphics[trim={0 80 230 30},width=0.7\linewidth]{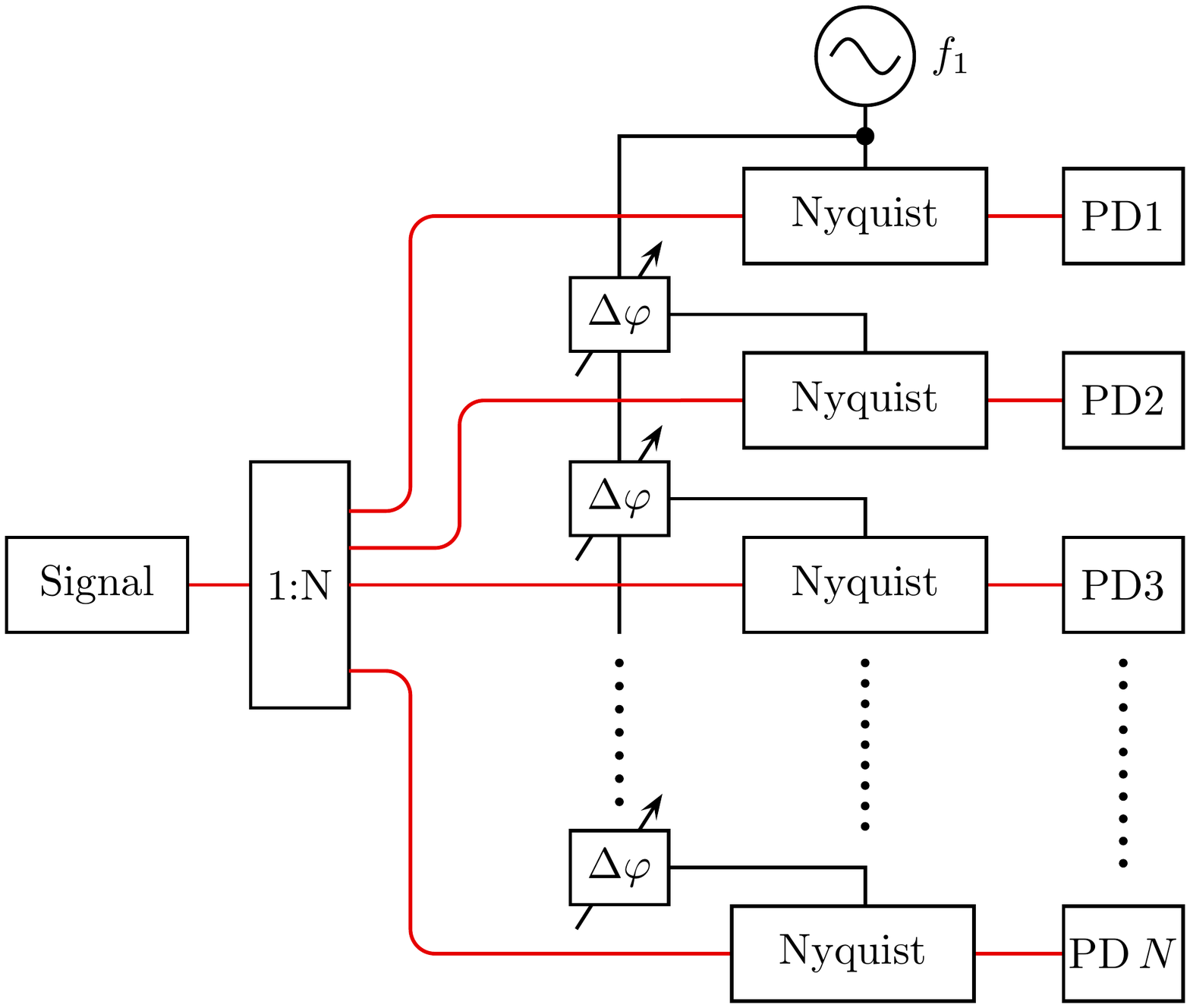}
\caption{Principle configuration for parallelisation in time domain. Input optical signals are fed to a $1:N$ coupler and transferred via optical fibres to $N$ contiguous Nyquist sequence branches, equally delayed by a phase shift of the electrical signal. Photodiodes (PD) are utilised for detection of the sampled optical signal. Of course, the setup can be changed in a way that the $N$ branches are fed with the Nyquist pulses generated from one source. Thereby, the multiplication by the signal can be accomplished by a modulator in each branch and a delay of the optical signal. Red lines are optical and black lines are electrical connections. The measurement was carried out with an OSO.}\label{fig:parall.time}
\end{figure}

The maximum achievable sampling rate for periodical signals is restricted to four times the modulator’s bandwidth. Besides, only one modulator driven with $4$ equidistant input frequencies can be used instead of two modulators (see the Methods for details).
For high-bandwidth non-periodic signals a parallelisation of the method is required. Then again the sampling rate can be defined by the bandwidth of the pulses $B=Nf_1$ and the sampling frequency is $f_s=B=1/\tau$. 
This parallelisation can be achieved in time or in frequency domain. In time domain, the optical input signal is coupled through a $1:N$ coupler into $N$ parallel branches as can be seen in Fig.~\ref{fig:parall.time}. In each branch the signal should be sampled at a fixed time delay compared to the previous. This time delay can be simply achieved by a phase change between the electrical sine waves, feeding the Nyquist sources. Thus, no optical tunable delays are required. Such tunable delay stages would only be possible by a mechanical tuning or by slow light methods \cite{filtering}.
\begin{figure}[!ht]
\centering
\includegraphics[trim={60 0 30 0},width=0.7\linewidth]{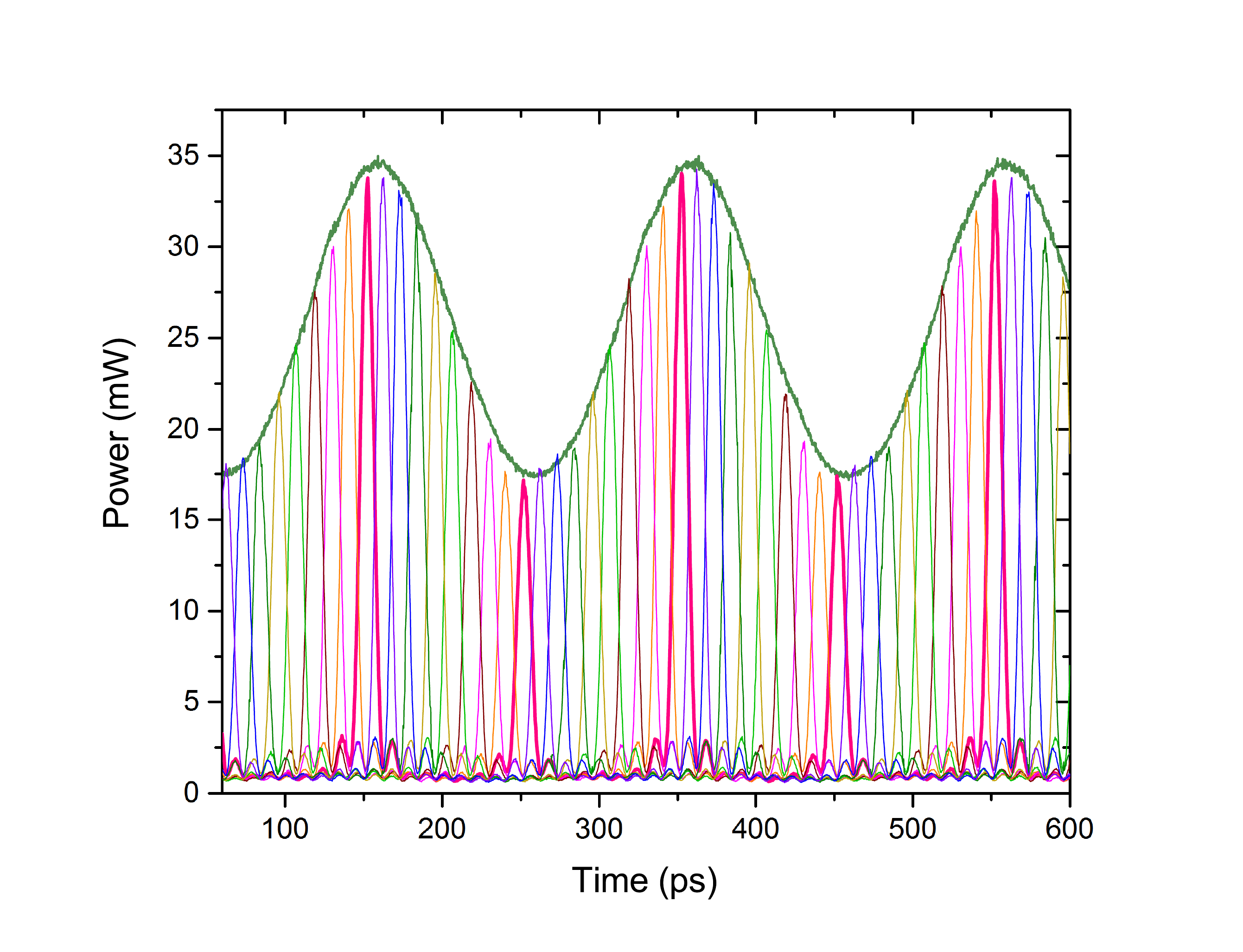}
\caption{Proof-of-concept parallelisation of optical sampling method. Sinusoidal waveform with frequency of $5$ GHz sampled by convolution in two coupled modulators. Each coloured trace represents a different sampling channel, here carried out by an electrical phase shift of $f_1$ and subsequent measurements.}\label{fig:parallel}
\end{figure}
As a substitute, a fast and simple adaptation of all sampling parameters to the signal is possible, since they are adjusted in the electrical domain. Due to parallelisation, the method grants $N$ times larger bandwidth with slow electronic components and photodiodes for non-periodic signals. \par
In order to demonstrate the parallelisation, in a proof-of-concept experiment the sampling points were time-shifted by an electrical phase shift of the frequency $f_1$. For every simulated channel a delay of $11$ ps is added equal to the pulsewidth $\tau$. Different measurements were carried out and plotted together in Fig.~\ref{fig:parallel}.\par
\begin{figure}[ht!]
\centering
\includegraphics[trim={40 300 10 100},width=0.7\linewidth]{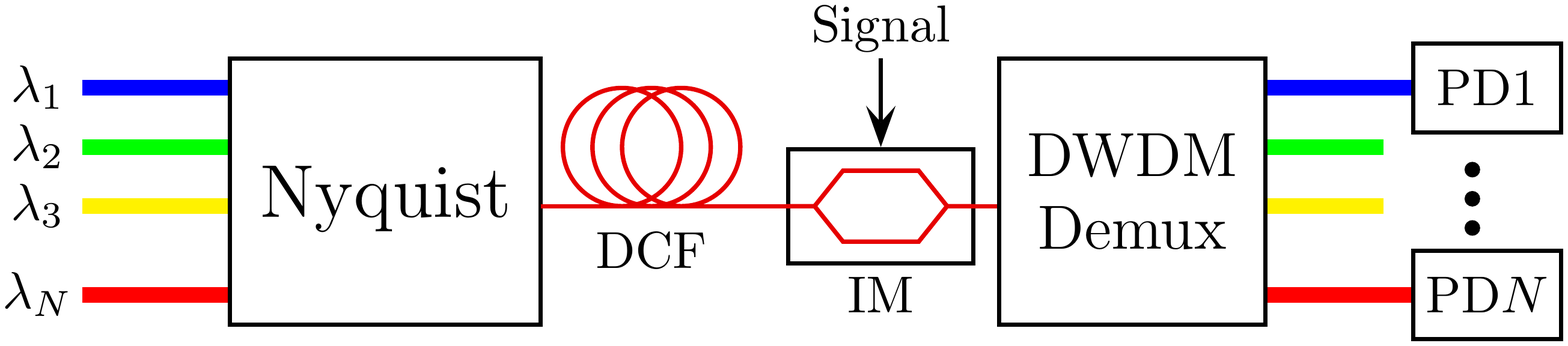}
\caption{Principle configuration for the parallelisation in frequency domain. Different input optical inputs with different wavelength ($\lambda_1$ to $\lambda_N$ ) are fed to a Nyquist sequence generating block (two coupled modulators driven with $f_1$ and $3f_1$, or one modulator driven with several coupled radio frequencies). The sequences are delayed in a dispersion compensating fibre (DCF). The multiplication in the time domain is obtained in an intensity modulator (IM). After demultiplexing of the sampled signals, the sampling points are detected by parallel photodiodes.}\label{fig:parall.freq}
\end{figure}

An alternative method, is to parallelise the sampling signal in frequency domain through several wavelengths as shown in Fig.~\ref{fig:parall.freq}. The approach is based on feeding a single Nyquist generator module (two coupled modulators, or one modulator driven with several coupled radio frequencies) with different input optical frequencies, causing generation of several frequency combs with central frequencies equal to the input optical frequencies. The following dispersion compensating fibre (DCF) delays the sinc-pulse sequences by $2\pi L\times k_2\times\Delta f$, where $L$ and $k_2$ are the length and group velocity dispersion of the DCF, respectively. $\Delta f$ is the frequency difference between the centre frequencies of the frequency combs. The delayed sequences are multiplied by the signal in an optical modulator. For higher sampling speeds, a nonlinear multiplication can be used. The sampling signals at different wavelengths are demultiplexed in an appropriate device and detected. 
In order to reduce costs, the frequency difference could correspond to a multiple of the ITU-T frequency grid for dense wavelength division multiplexing ($n\times50$ GHz) \cite{ITU-T}. Thus, CW lasers from optical telecommunications can be used as sources and the demultiplexing can be performed with an arrayed waveguide grating for instance. The maximum achievable sampling frequency of the signal, or the sampling rate of the system, increases with the number of wavelengths, $N$.
During parallelisation, the bandwidth requirements for the electronics remain the same for both methods. This means, that low bandwidth electronics can sample non-periodic signals with a bandwidth that is increased by $N$.\par

In principle this is similar to the time-parallelisation approach, where the sampling rate is increased with the number of parallel branches. Here the time delay between the sampling points is controlled by changing the frequency difference between the optical input waves.\par
\subsection*{Conclusion}
In conclusion, a method for tunable all-optical sampling without requiring a pulse source based on time-frequency coherence was presented. Two coupled modulators were used to convolve the spectrum of the optical input signal with a rectangular frequency comb, corresponding to a multiplication between the time domain representation of the signal and a sinc-pulse sequence. The parameters of the sampling points (sampling rate, bandwidth, position) are just defined by the sinusoidal input used to drive the modulators. Thus, they can be fast and easily adapted to the signal and a time shift is possible by a phase change. The maximum bandwidth of the pulses, or their minimum duration and therefore the sampling rate is restricted by the bandwidth of the modulators. With two intensity modulators, the maximum achievable bandwidth is $4f_1$ (see the Methods for details). Thus, with integrated $100$ GHz modulators \cite{100GHz}, a sampling rate of $400$ Gs/s is possible. Sampling rates in the Tb/s range can be achieved by modulators driven with phase-locked optical frequencies \cite{Flexible} and by a balanced detector the full field can be sampled \cite{LinOs}.
\section*{Supplementary Information} 
\subsection*{Sampling}
\begin{figure}[!ht]
\centering
\includegraphics[width=\linewidth]{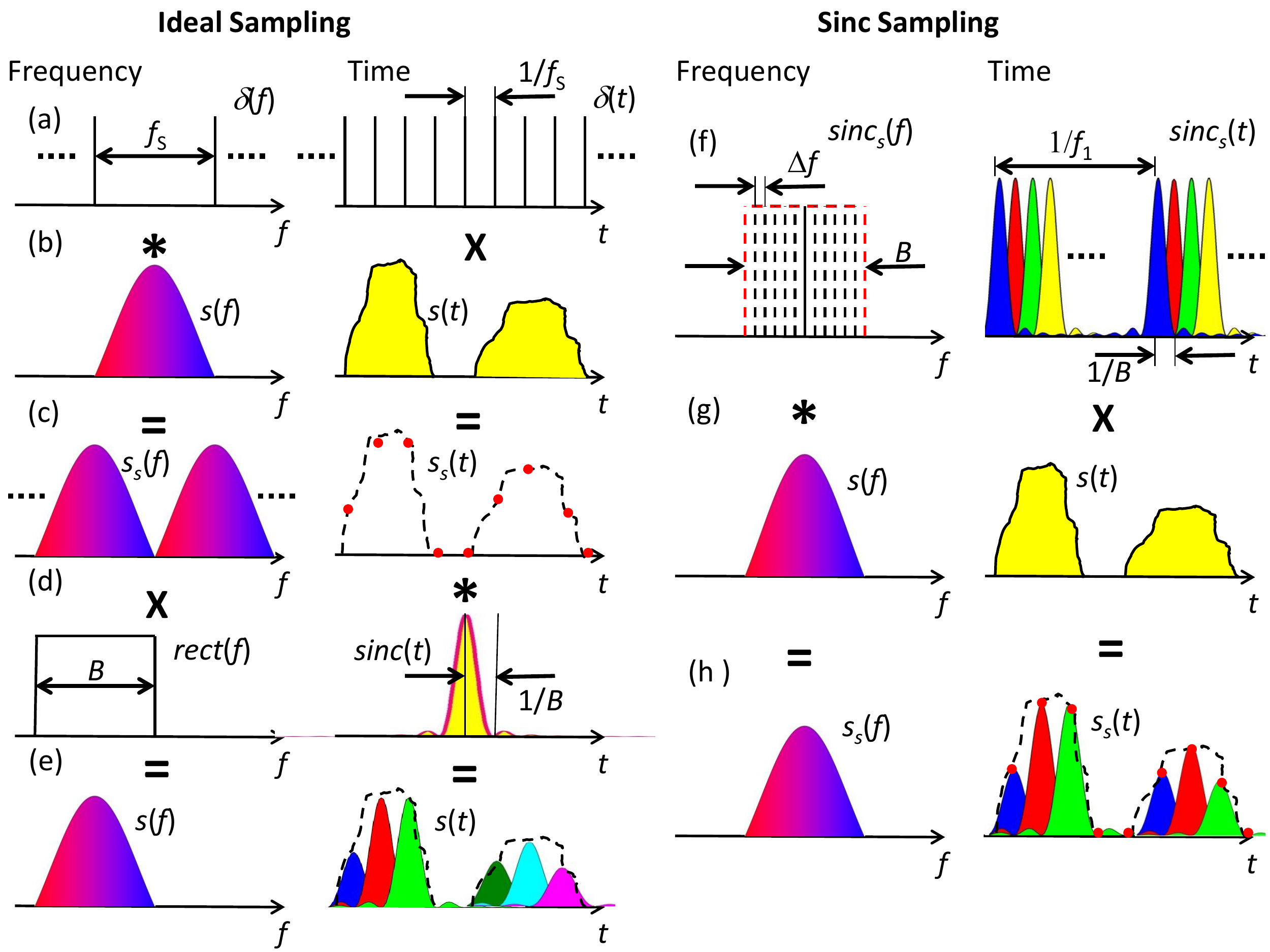}
\caption{Sampling of a signal in the frequency and time domain with a Dirac-Delta function (left), corresponding to \enquote{ideal} sampling and a sinc-pulse sequence (right). The $\times$ and $\ast$ denote the signs for a multiplication and convolution, respectively. The $=$ marks the corresponding result of the operation, sincs(t) is the sinc-pulse sequence defined in equation~\eqref{Nyquist}}\label{fig:principle}
\end{figure}
To convert a continuous to a digital signal in time domain, a sampling of the time domain waveform is required. The analogue signal to sample $s(t)$ (Fig.~\ref{fig:principle}b) is supposed to be a band limited function with a baseband bandwidth of $B_s$ and the signal can be recovered, if the sampling rate (number of samples per second) is at least twice the maximum frequency present in the signal spectrum $f_s=1/t_s>2B_s$. Here $t_s$ is the time duration between two samples, or the inverse of the sampling rate (Fig.~\ref{fig:principle}a, right). The sampled discrete signal in time domain is (Fig.~\ref{fig:principle}c, right):
\begin{align}
\label{sampling time}
s_s(t)=s(t)\cdot\sum^{\infty}_{n=-\infty}\delta\bigg(t-\frac{n}{f_s}\bigg)\nonumber\\
=\sum_{n=-\infty}^{\infty}s\bigg(\frac{n}{f_s}\bigg)\delta\bigg(t-\frac{n}{f_s}\bigg),
\end{align}
where $\delta(t)$ is an ideal Dirac delta function. In the frequency domain this corresponds to a convolution ($\ast$) between the frequency domain representations of the signal $S(f)$ and the Dirac delta function (Fig.~\ref{fig:principle}a, left):
\begin{align}
\label{sampling frequency}
S_s(f)=S(f)\ast\sum_{n=-\infty}^{\infty}\delta(f-nf_s)\nonumber\\
=\sum_{n=-\infty}^{\infty}S(f-nf_s).
\end{align}
The result is an unlimited number of copies of the signal spectrum with a frequency distance of $f_s$. This sampled signal can be reconstructed by a multiplication with a rectangular function in the frequency domain (Fig.~\ref{fig:principle}d, left). This multiplication filters just one of the unlimited number of spectra. 
\begin{equation}\label{S(f)}
S(f)=\sum_{n=-\infty}^{\infty}S(f-nf_s)\cdot \text{rect}\bigg(\frac{f}{f_s}\bigg)
\end{equation}
The result is the analogue signal (Fig.~\ref{fig:principle}e, right): 
\begin{eqnarray}\label{s(t)}
s(t)&=\Bigg[\sum_{n=-\infty}^{\infty}s\bigg(\frac{n}{f_s}\bigg)\delta\bigg(t-\frac{n}{f_s}\bigg)\Bigg]\ast\text{sinc}(f_s t)\nonumber\\
&=\sum_{n=-\infty}^{\infty}s\bigg(\frac{n}{f_s}\bigg)\text{sinc}\Bigg(f_s\bigg(t-\frac{n}{f_s}\bigg)\Bigg),
\end{eqnarray}
with sinc as the sinus cardinals or sinc-function as defined above. Thus, a band limited analogue signal can be seen as a superposition of an unlimited number of sinc-pulses weighted with the sampling value and time shifted by the sampling time.\par
The sampling with the method presented here leads in the frequency domain to (Fig.~\ref{fig:principle}h, left):
\begin{equation}
\label{Ss(f)}
S_s(f)=S(f)\ast\Bigg[\sum_{n=-\infty}^{\infty}\delta(f_c-nf_s)\cdot\text{rect}\bigg(\frac{f-f_c}{B}\bigg)\Bigg]
\end{equation}
with $f_c$ as the carrier frequency of the optical wave. Thus, the copies of the spectrum are not unlimited as for the sampling with a Dirac Delta function, but restricted to the bandwidth $B=Nf_1=Nf_s$. For a parallelised sampling it can be written that $B=f_1=f_s$. Therefore, the sampled signal corresponds to just one copy of the spectrum. Thus, in time domain it follows (Fig.~\ref{fig:principle}h, right):
\begin{eqnarray}\label{ss(t)}
s_s(t)=&s(t)\cdot\Big[\sum_{n=-\infty}^{\infty}\delta(t-\frac{n}{f_s})\ast\text{sinc}(Bt)\Big]\nonumber\\
=&s(t)\cdot\sum_{n=-\infty}^{\infty}\text{sinc}\big(B(t-\frac{n}{f_s})\big)\nonumber\\
\approx&\sum_{n=-\infty}^{\infty}s\big(\frac{n}{f_s}\big)\text{sinc}\big(B(t-\frac{n}{f_s})\big).
\end{eqnarray}
By the comparison with equation~\eqref{s(t)} it can be seen that the method grants access to the sampling values $s(n/f_s)$ by an integration of the sinc-pulse sequence over the time interval $1/f_1$. This integration can be carried out with a photodiode with a restricted bandwidth of $f_1=B/N$.\par
\subsection*{Tunability}
According to the left hand side of equation~\eqref{f-1}, the frequency representation of the sinc-pulse sequence is a rectangular frequency comb. The individual comb lines have to have the same or a linear dependent phase. A simple method to generate such a frequency comb is to use two coupled intensity modulators. Both modulators are adjusted in a way that the first order sidebands have the same amplitude as the carrier and higher order sidebands are suppressed. If the DC bias and the RF signal voltages applied to a single modulator are $V_B$ and $v_s\cos(\omega_st)$, respectively, the expression for the output field can be written as \cite{Bessel}
\begin{figure}[ht!]
\label{f1f2}
\centering
\begin{subfigure}[b]{0.3\textwidth}
\centering
\includegraphics[trim={30 30 20 30},width=\textwidth]{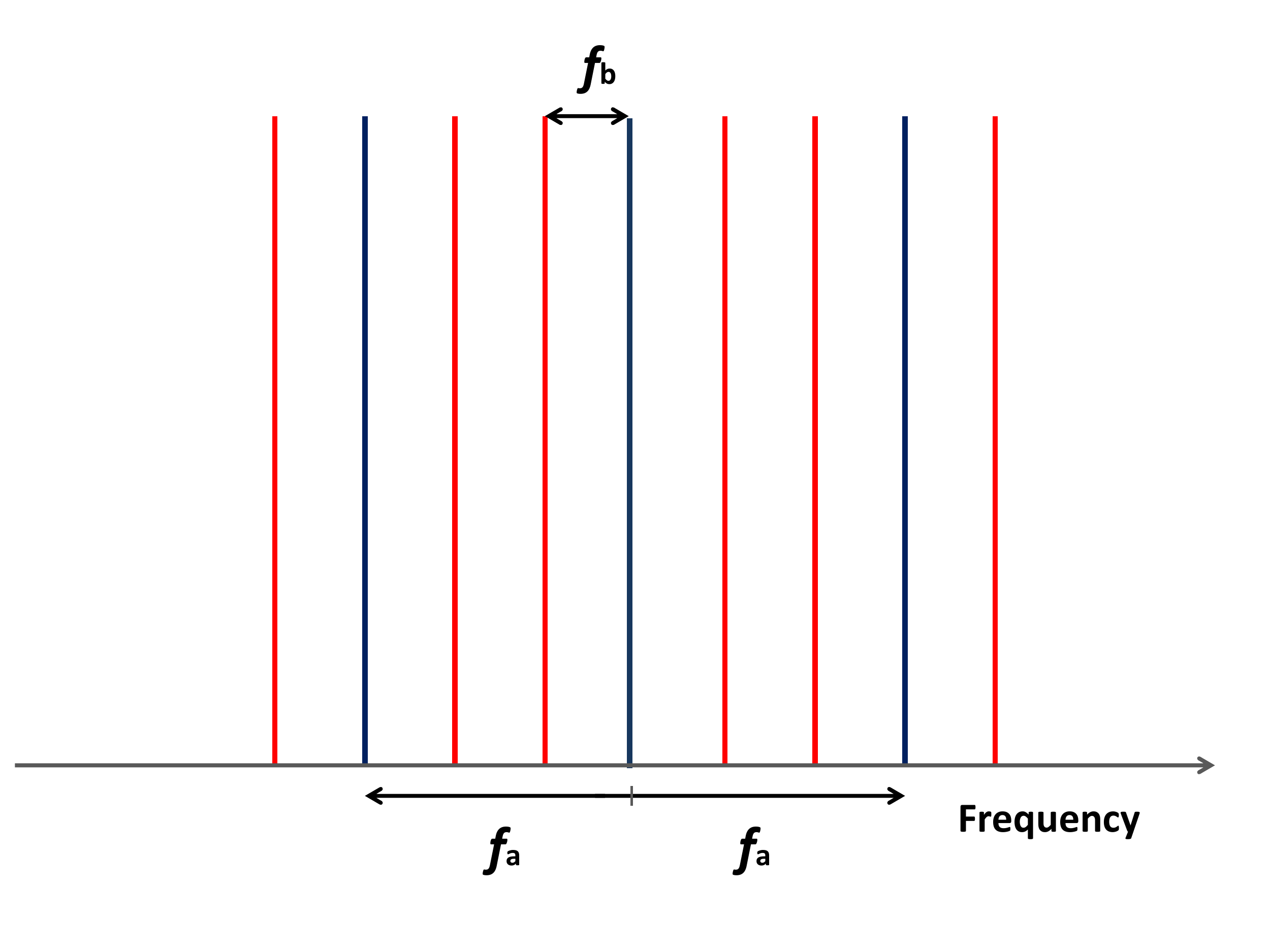}
\caption{}\label{fig:2a}
\end{subfigure}
\centering
\begin{subfigure}[b]{0.3\textwidth}
\centering
\includegraphics[trim={30 30 20 30},width=\textwidth]{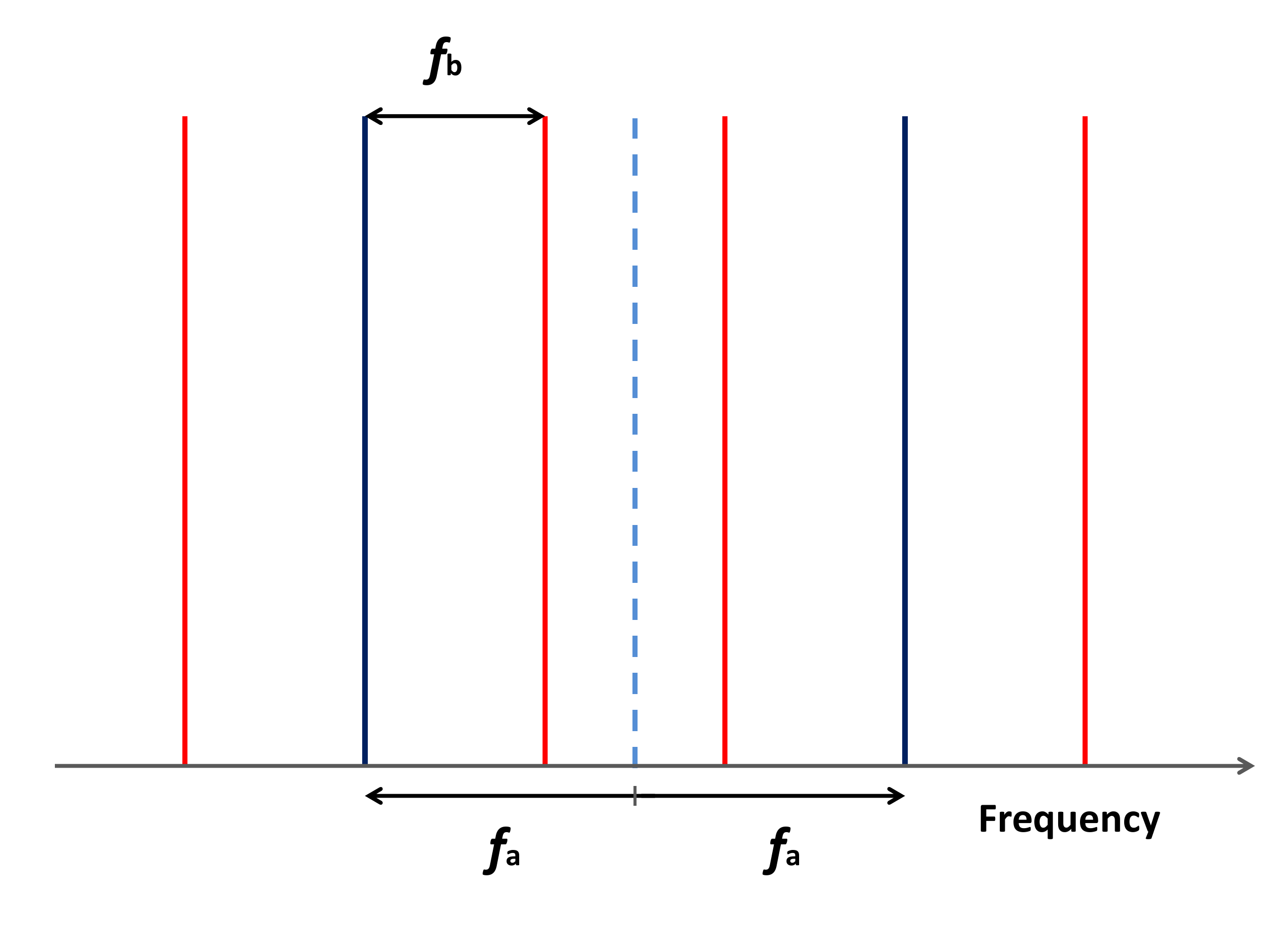}
\caption{}\label{fig:2b}
\end{subfigure}
\centering
\begin{subfigure}[b]{0.3\textwidth}
\centering
\includegraphics[trim={30 30 20 30},width=\textwidth]{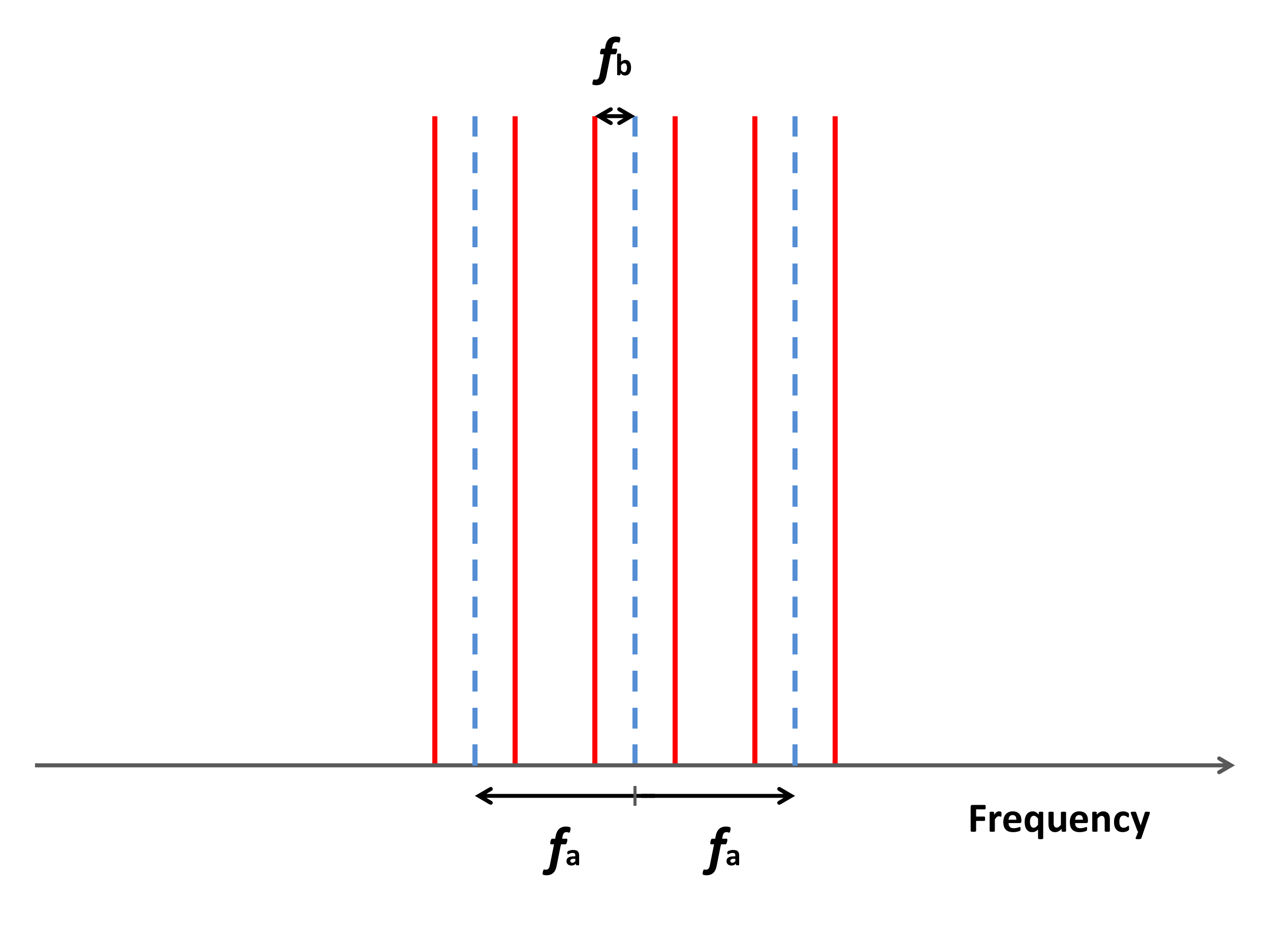}
\caption{}\label{fig:2c}
\end{subfigure}
\caption{(a) 9-line frequency comb, the $3$ blue spectral lines are generated by the first modulator, whereas the second modulator generates the red lines; (b) Frequency comb with $6$ lines. The first modulator (blue lines) operates in the suppressed carrier regime (dashed blue line) and the second modulator generates the red lines; (c) Both modulators operate in the suppressed carrier regime (dashed lines). The result is again a $6$ line frequency comb.}\label{fig:f1f2}
\end{figure}
\begin{equation}
\label{Bessel_1}
\begin{split}
E(t)=\sum_{n=-\infty}^{\infty}(-1)^{k}\big\{\cos(\pi\epsilon)J_{2k}(\pi\alpha)\cos(\omega_0 t+2k\omega_st)\\
+\sin(\pi\epsilon)J_{2k-1}(\pi\alpha)\cos(\omega_0 t+(2k-1)\omega_st)\big\},
\end{split}
\end{equation}
where $J_k$ is the Bessel function of the first kind and order $k$, $\epsilon=V_B/V_{\pi}$ and $\alpha=v_s/V_{\pi}$, in which $V_{\pi}$ is the half-wave voltage of the modulator. Contrary to phase modulators, the amplitude of the carrier, first-order sidebands and higher-order sidebands in intensity modulators can be adjusted by two parameters, i.e. the RF driving voltage $\alpha$ and the DC bias $\epsilon$. If equation~\eqref{Bessel_1} is reduced to the spectral components of interest, i.e. the carrier and the first order sidebands, the electric field at the output of the modulator can be written as \cite{Nature}:
\begin{equation}
\label{Bessel_2}
\begin{split}
E(t)=\cos(\pi\epsilon)J_0(\pi\alpha)\cos(\omega_0 t)-\sin(\pi\epsilon)J_1(\pi\alpha)\\
\big\{\cos((\omega_0-\omega_s)t)+\cos((\omega_0+\omega_s)t)\big\}.
\end{split}
\end{equation}
If two intensity modulators are used and both have just one input frequency, then there are three possibilities of generating a rectangular frequency comb as illustrated in Fig.~\ref{fig:f1f2}.\par
If the carrier is not suppressed as in Fig.~\ref{fig:2a}, the modulators generate $9$ lines. The first modulator is driven with the frequency $f_a$. Thus, the modulator generates the three blue lines. In the second modulator these lines are re-modulated with the frequency $f_b=\frac{1}{3}f_a$ to produce the red lines. If at least one of the modulators can be driven with a maximum frequency of $f_a=40$ GHz, the maximum bandwidth is $B=N\times f_1=N\times f_b=120$ GHz, which corresponds to a pulse width (FWHM) of around $8$ ps. In Fig.~\ref{fig:2b} the carrier of the first modulator is suppressed. Thus, an even number of lines ($6$) is produced. Here the second modulator has to be driven with $f_b=\frac{2}{3}f_a$ and the maximum bandwidth is $N\times f_b=160$ GHz, corresponding to $6$ ps pulses. If both modulators are driven in the suppressed carrier regime, again an even number of lines is produced but now with a maximum bandwidth of $B=60$ GHz ($17$ ps) since $f_b=\frac{1}{4}f_a$. Of course the first modulator can be driven with the lower and the second with the higher frequency, however the results are the same.\par
The repetition period of the Nyquist pulses, and therefore the number of parallel branches or the wavelength for sampling, is defined by the number of lines. Nine lines as in Fig.~\ref{fig:2a} can be produced if just one modulator is driven with a four-frequency input signal.
Thus, each odd number $N$ of lines can be produced with just one modulator if the RF-signal driving the modulator consists of $(N-1)/2$ equally spaced frequency lines, for an even number of lines the number of RF is $N/2$. 
These equally spaced frequency lines in the RF domain can be produced by a mixing of higher harmonics, or by an arbitrary waveform generator for instance. However, the maximum bandwidth of this method is restricted by around twice the bandwidth of the used modulator. Another possibility to increase the number of lines is the usage of more than two modulators. However, the maximum bandwidth which can be achieved does not increase much with the number of additional modulators. Higher bandwidths can be achieved if higher order sidebands are used. Although, since the carrier and the first order sidebands are out of phase, this requires that these components are effectively suppressed. One possibility to achieve Nyquist pulses with very broad bandwidths is the filtering of the frequency comb of a mode-locked source \cite{filtering}. However, due to the non-ideal bandwidth of optical filters, the generated pulses have no sinc shape and the pulse parameters cannot be adapted easily to the sampled signal. Instead of filtering a whole comb as rectangular as possible, single comb lines can be filtered. These lines are the initial lines for a subsequent modulation. The initial lines are responsible for the arbitrary bandwidth, not limited by modulators, and the subsequent modulation enables an arbitrary repetition period of the pulses. As a mode-locked source, a fibre fs-laser or much cheaper and simpler, a comb generator can be used. In proof-of-concept experiments, sinc-shaped Nyquist pulses with a bandwidth of around $300$ GHz have been generated by this method \cite{Flexible}. However, bandwidths in the THz-range and correspondingly femtosecond pulses are possible from a comb generator.\par
\section*{Acknowledgements}
We acknowledge the funding from the Deutsche Forschungs Gemeinschaft (DFG) (SCHN716/13-1).\\
%
\section*{Author contributions}
S.P. and G.R.M. performed the experiments. T.S. proposed the first concept of the optical source free sampling and the theory. All authors contributed equally in writing the manuscript.
\section*{Additional information}
\textbf{Competing financial interests:} The authors declare that they have no competing financial interests.\\

\end{document}